\documentclass[aps,prl
,twocolumn,amssymb,amsmath,a4paper,superscriptaddress,longbibliography]{revtex4-2}
\usepackage{graphicx}
\usepackage{amsfonts}
\usepackage{amsmath}
\usepackage{hyperref}
\usepackage{xr-hyper}
\usepackage{mathrsfs}
\usepackage{adjustbox}
\usepackage{float} 
\usepackage[normalem]{ulem}
\usepackage[nameinlink]{cleveref}
\usepackage[table,xcdraw]{xcolor}
\usepackage[mathcal]{euscript}
\Crefname{figure}{Fig.}{Figs.}
\def\be{\begin{eqnarray}}
\def\ee{\end{eqnarray}}
\def\nn{\nonumber}

\begin{document}

\title{Gravitational Mode Instabilities in Kerr-like Black Holes}

\author{Pedro Henrique Croti Siqueira}
\email{pedro.croti@ufabc.edu.br}
\affiliation{Centro de Matem\'atica, Computa\c c\~ao e Cogni\c c\~ao, Universidade Federal do ABC (UFABC), 09210-170 Santo Andr\'e, S\~ao Paulo, Brazil}
\affiliation{Centro de Ci\^encias Naturais e Humanas, Universidade Federal do ABC (UFABC), 09210-170 Santo Andr\'e, S\~ao Paulo, Brazil}
\affiliation{Instituto de Física Gleb Wataghin, Universidade Estadual de Campinas, 13083-059 Campinas, São Paulo, Brazil}
\author{Rodrigo Dal Bosco Fontana}
\email{rodrigo.dalbosco@ufrgs.br}
\affiliation{Universidade Federal do Rio Grande do Sul (UFRGS), 95590-000, Tramandaí, Rio Grande do Sul, Brazil}

\author{Maur\'icio Richartz}
\email{mauricio.richartz@ufabc.edu.br}
\affiliation{Centro de Matem\'atica, Computa\c c\~ao e Cogni\c c\~ao, Universidade Federal do ABC (UFABC), 09210-170 Santo Andr\'e, S\~ao Paulo, Brazil}

\begin{abstract}
\noindent 
We identify exponentially growing modes of a homogeneous gravitational Teukolsky-like operator on highly spinning Kerr-like black holes subject to quasinormal mode boundary conditions. Specifically, we compute electromagnetic and gravitational quasinormal modes of a spacetime proposed by Konoplya, Rezzolla, and Zhidenko that parametrizes deformations from the Kerr geometry, including configurations with black hole spin parameter $a/M > 1$ that retain an event horizon. Using the continued fraction method, we find that the spin-weight $s=-2$ operator develops exponentially growing modes when the black hole spin and  deformation parameters are sufficiently large, while the electromagnetic modes remain damped
throughout the parameter space. We further compare the quasinormal mode predictions with an agnostic analysis of the GW250114 ringdown, identifying regions of the parameter space that are compatible with the spectrum inferred from the data.

\end{abstract}
\textcolor{red}{}
\maketitle
%%%%%%%%%%%%%%%%%%%%%%%%%%%%%%%%%%%%%%%%%%%%%%%%%%%%%%%%%%%%%%%%%%%%%%%%%%%%%%%%%%%%%%%%%%%%%%%%

\noindent \textbf{\textit{Introduction --}} The Kerr spacetime
is the unique stationary, axisymmetric, vacuum black hole solution of General Relativity (GR), specified by its mass $M$ and spin $a$~\cite{Carter:1971zc, Robinson:1975, Cardoso:2016ryw}. If astrophysical black holes reach stationarity
and are well approximated as vacuum objects, they are expected to be described by the Kerr metric.
This expectation is encapsulated in the Kerr hypothesis, which states that all astrophysical black holes are described by the Kerr geometry~\cite{Bambi:2011mj,Herdeiro:2022yle}. Gravitational wave (GW) observations of compact binary coalescences 
offer increasingly precise tests of this hypothesis, using both the inspiral and ringdown phases~\cite{LIGOScientific:2026oim, Isi:2019aib, Brito:2018rfr,Cardenas-Avendano:2019zxd,Berti:2025hly}
to probe deviations from Kerr. In this context, one phenomenological strategy is to constrain parametrically deformed Kerr-like geometries using observational data.

The Konoplya--Rezzolla--Zhidenko (KRZ) framework parametrizes stationary, axisymmetric deviations from Kerr~\cite{Konoplya:2016jvv}. X-ray spectroscopy and black hole shadow observations probe these deformations, although their constraints are affected by parameter degeneracies and modeling uncertainties~\cite{Abdikamalov:2021zwv,Yu:2021xen,Bambi:2016iip,Olmo:2025ctf}. GW observations provide complementary constraints through inspiral and ringdown signals~\cite{Cardenas-Avendano:2019zxd,Konoplya:2016pmh,Shashank:2021giy}. Inspiral analyses evaluate metric deformation corrections in the nonrotating limit, neglecting spin-deformation couplings~\cite{Cardenas-Avendano:2019zxd,Shashank:2021giy}, while single mode ringdown studies suffer from parameter degeneracies~\cite{Cardenas-Avendano:2016zml,Konoplya:2016pmh}.
Recently, data from GW250114 were used to constrain KRZ deformations in the regime $a<M$ through eikonal quasinormal modes (QNMs)~\cite{DeSimone:2026jqv}.
 This approach, however, does not capture the full spin-dependent perturbation dynamics (see also \cite{Volkel:2026qqz}).

More fundamentally, specifying a parametrized black hole geometry is not sufficient to determine the gravitational perturbation equations, which also depend on the underlying gravitational theory. An alternative theory-agnostic strategy is to characterize the perturbation operators and their possible couplings directly~\cite{Volkel:2022aca,Cano:2024jkd}. Such operators need not be stable: growing modes have been reported in different parametrized models~\cite{deMedeiros:2025ayq,DeSimone:2026waz}. Gravitational instabilities have also been identified in horizonless, superspinning ($a>M$) Kerr objects, for which an inner boundary must be prescribed, as well as in higher-dimensional rotating black holes~\cite{Pani:2010jz,Santos:2026pbu}. These results motivate us to investigate whether a homogeneous gravitational Teukolsky-like operator, subject to QNM boundary conditions, can admit growing modes on a four-dimensional Kerr-like black hole.

Here we compute the QNM spectra of test electromagnetic perturbations and gravitational perturbations governed by a homogeneous spin-weight $s=-2$ Teukolsky-like operator on a one-parameter subclass of the KRZ spacetime that includes superspinning configurations with an event horizon. We identify exponentially growing $(\ell,m)=(2,2)$ and $(2,1)$ modes of the gravitational operator at sufficiently large black hole spins and deformation parameters, while the electromagnetic modes remain damped. To the best of our knowledge, these are the first growing modes identified in the spectrum of a homogeneous spin-weight \(s=-2\) Teukolsky-like operator on four-dimensional Kerr-like black holes subject to QNM boundary conditions. In addition to studying the perturbation spectra and the instabilities, we 
also provide a phenomenological test of the Kerr hypothesis by identifying regions of the parameter space that yield frequencies compatible with those inferred from an agnostic analysis of the GW250114 ringdown~\cite{LIGOScientific:2025wao,LIGOScientific:2025rid}.

Throughout this work, we use geometrical units ($G = c = 1$) unless otherwise stated.

\noindent \textbf{\textit{Kerr-like black holes --}} We consider a stationary and axisymmetric Kerr-like spacetime, described in Boyer–Lindquist coordinates $(t, r, \theta, \phi)$ by the line element
\begin{align}
	ds^{2} = &-\frac{\Delta-W^{2} r^2 \sin ^{2}\theta }{K^{2}}dt^{2} - 2Wr\sin ^{2}\theta \, dt d\phi \nonumber \\
	&+ K^{2}\sin ^{2}\theta \, d\phi ^{2}+\Sigma \left( \frac{dr^{2}}{\Delta}+d\theta ^{2}\right), \label{eq:metric}
\end{align}
where
\begin{subequations}
\begin{align}
&	\hspace*{-0.25cm} \Sigma = r^{2}+a^{2}\cos ^{2}\theta, \quad K^{2} = \frac{\left( r^{2}+a^{2}\right) ^{2}-a^{2}\Delta\sin ^{2}\theta}{\Sigma}, \\
&	\hspace*{-0.25cm} W = \frac{a(2Mr^2 + \eta)}{r^2 \Sigma }, \quad \,  \Delta = r^{2}-2Mr+a^{2}-\frac{\eta}{r}.  \label{eq:Delta}
\end{align}
\end{subequations}
The constant $\eta$ parametrizes deviations from the Kerr geometry, which is recovered when $\eta = 0$. The three roots $(r_0,r_1,r_2)$ of $\Delta$ satisfy $\sum_{i} r_i = 2M$, $\prod_{i} r_i = \eta$, and $\sum_{i<j} r_i r_j = a^2$. The largest positive real root, $r_0$, defines the event horizon, with angular velocity $\Omega_0=a/(r_0^2+a^2)$ (see Fig.~\ref{fig:roots}). While superspinning Kerr spacetimes correspond to naked singularities, a superspinning configuration of the Kerr-like geometry \eqref{eq:metric} may represent a black hole if the deformation parameter $\eta$ is sufficiently large. The Kerr-like geometry also displays an ergoregion, which resembles that of Kerr for small deformation parameters but can fragment into disconnected pieces when $a >M$~\cite{Franzin:2021kvj}.
\begin{figure}[htpb]
    \centering   \includegraphics[width=0.99\linewidth]{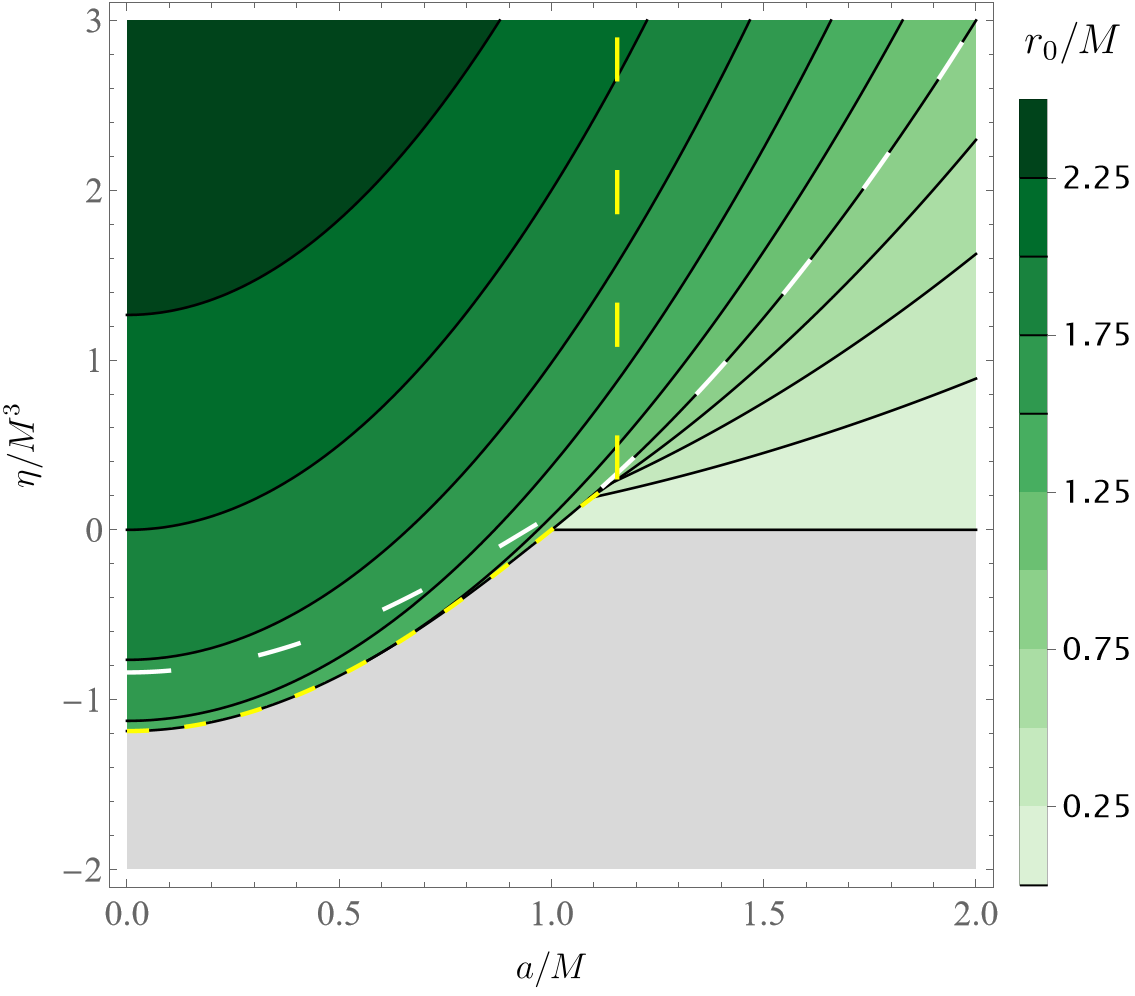}
    \caption{The location $r_0/M$ of the event horizon of the Kerr-like black hole \eqref{eq:metric}. The gray region corresponds to naked singularities. The dashed yellow and white lines delimit the regions where the continued fraction method is applicable (see Supp. Material~\cite{supplemental}).
}
    \label{fig:roots}
\end{figure}

\noindent \textbf{\textit{Electromagnetic and gravitational perturbations --}} We treat the metric \eqref{eq:metric} as a phenomenological Kerr-like template, and study linear perturbations within the Newman-Penrose formalism. In GR, the separability of perturbations on a Kerr background relies on the spacetime being both Petrov type-D and Ricci-flat~\cite{Cano:2023tmv}. While our deformed metric preserves the Petrov type-D classification, it is not Ricci-flat, exhibiting non-vanishing background Ricci scalars. These curvature terms couple the perturbation equations and prevent the straightforward decoupling of the gravitational variables (as in the Kerr--Newman spacetime, where separating the coupled gravitoelectric system has recently been achieved using a matrix integrating factor~\cite{Hintz:2026jiw,Destounis:2026ljd}). Following the separable-gauge construction for non-vacuum type-D spacetimes \cite{guo24,nakajima25}, we consider the homogeneous sector of a Teukolsky-like equation obtained by imposing a suitable tetrad gauge condition and setting the effective source term to zero (see Supp. Material~\cite{supplemental}). The resulting equation defines a separable eigenvalue problem. Scalar and electromagnetic perturbations, in contrast, decouple and separate directly~\cite{Franzin:2021kvj}.

\begin{figure*}[!htbp]
	\centering
	\includegraphics[width = 1.99 \columnwidth]{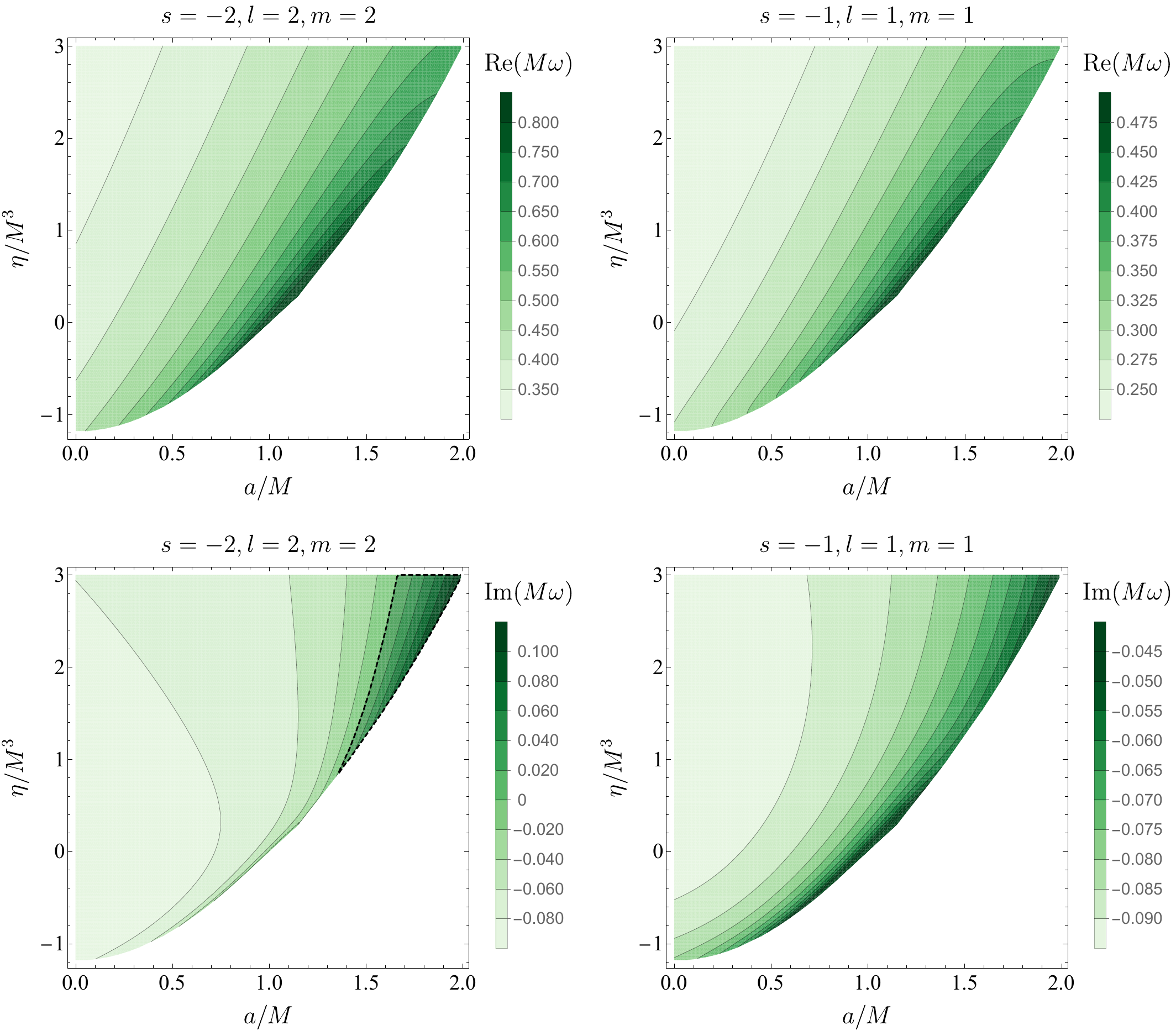}
	\caption{Contour plots of the fundamental QNM frequencies in the $(a/M, \eta/M^3)$ parameter space.  The left and right columns show $(\ell,m)=(2,2)$ gravitational and $(\ell,m)=(1,1)$ electromagnetic perturbations, respectively.
    The top panels show the oscillation frequency $\mathrm{Re}(M\omega)$, while the bottom panels show the imaginary part $\mathrm{Im}(M\omega)$. The black dashed lines in the bottom-left panel enclose the region where $\operatorname{Im}(M\omega)>0$, indicating the presence of exponentially growing modes of the homogeneous spin-2 operator.}
	\label{fig:s2es1} 	
\end{figure*}

In all cases, we employ the standard \textit{ansatz} $\psi_s = e^{-i\omega t + im\phi} R_s(r) S_s(\theta)$, with $s=0,-1,-2$ denoting scalar, electromagnetic, and gravitational perturbations, respectively. This decomposes the spin-s field $\psi_s$ into modes of frequency $\omega$ and azimuthal number $m$, separating the equations of motion into angular and radial parts. The angular functions $S_s(\theta)$ are spin-weighted spheroidal harmonics with separation constant $\Lambda_{\ell m}$ indexed by the orbital number $\ell$~\cite{Berti:2005gp}.
The dynamics of the perturbations are governed by the following master equation for the radial function $R_s(r)$ (see Supp. Material~\cite{supplemental}):
\begin{equation} \label{eq:master_eq}
\Delta^{-s} \frac{d}{dr} \left( \Delta^{s+1} \frac{dR_s}{dr} \right) + V_s(r) R_s(r) = 0,
\end{equation}
where 
\begin{equation} \label{eq:potential}
V_s(r) = \frac{\mathcal{K}^2 - i s \mathcal{K} \Delta'}{\Delta} + 4 i s \omega r + 2am\omega - a^2\omega^2 - \Lambda_{lm},
\end{equation}
primes denote derivatives with respect to $r$, and $\mathcal{K} = (r^2 + a^2)\omega - am$.
The master equation \eqref{eq:master_eq} and the associated potential \eqref{eq:potential} have essentially the same form as the standard radial Teukolsky equation~\cite{tkk1,tkk2,tkk3},
with deviations from Kerr encoded in the function $\Delta$. 
Such deviations modify the near horizon potential and, therefore, affect the strong field regime and the QNM spectrum. Replacing $\Delta$ with $\Delta_{\rm KN}=r^2-2Mr+a^2+Q^2$, where $Q$ is the black hole charge, recovers the Dudley--Finley radial equation for Kerr--Newman perturbations~\cite{Dudley:1977zz,Dudley:1978vd}. This correspondence provides a useful consistency check for our separable formulation (see Supp. Material~\cite{supplemental}).
%(see Supp. Material~\cite{supplemental}, Sec.~III).

To determine the QNM spectrum we impose boundary conditions corresponding to purely ingoing waves at the horizon and purely outgoing waves at spatial infinity. Near the horizon, the radial solution behaves as $R_s(r) \sim \Delta^{-s} e^{-i k_0 r_*}$, where $r_*$ is the tortoise coordinate defined by $dr_*/dr = (r^2+a^2)/\Delta$ and $k_0 = \omega - m\Omega_0$ represents the horizon wavenumber. In the asymptotic region, the purely outgoing solution behaves as
$R_s(r) \sim r^{-2s-1} e^{+ i \omega r_*}$,
as required by the QNM boundary condition. We determine the discrete spectrum of  QNM frequencies satisfying these boundary conditions using Leaver's continued fraction method~\cite{Leaver:1985ax} (see Supp. Material~\cite{supplemental}).

\noindent \textbf{\textit{Numerical results --}}  We compute QNMs throughout the parameter space spanned by the dimensionless spin $a/M$ and deformation parameter $\eta/M^3$, covering the region accessible to the continued fraction method (see Fig.~\ref{fig:roots}). We focus on the dominant perturbations, namely the quadrupole $(\ell,m)=(2,2)$ mode in the gravitational sector and the dipole $(\ell,m)=(1,1)$ mode in the electromagnetic sector. Alongside the calculation of the standard damped spectra, with $\mathrm{Im}(M\omega)<0$, we also search for mode instabilities characterized by exponentially growing perturbations with $\mathrm{Im}(M\omega)>0$. 

The results are shown in Fig.~\ref{fig:s2es1}. Across the explored parameter space, the dominant gravitational mode oscillates at a higher frequency than the dominant electromagnetic mode for any given pair $(a/M,\eta/M^3)$. The comparison between their damping rates, on the other hand, is less straightforward. The gravitational mode has a larger $\mathrm{Im}(M\omega)$ than the electromagnetic mode in most of the parameter space, indicating that gravitational perturbations typically decay more slowly. The difference becomes particularly significant at large spins and deformations, where the two sectors exhibit qualitatively distinct behaviour. For electromagnetic perturbations, $\mathrm{Im}(M\omega)$ remains negative throughout the explored parameter space, including the superspinning regime, so the modes are always damped. In contrast, as shown in the bottom-left panel of Fig.~\ref{fig:s2es1}, the spin-$2$ operator develops exponentially growing modes, characterized by $\operatorname{Im}(M\omega)>0$, for sufficiently large black hole spins and sufficiently large deformations.

We quantify this gravitational instability for two representative configurations. Near the boundary of the unstable region, at $(a/M,\eta/M^3)=(1.36,0.85)$, we find $\mathrm{Im}(M\omega)\simeq3.4\times10^{-3}$, corresponding to an instability timescale $\tau\simeq2.9\times10^2M$, or approximately $93\,\mathrm{ms}$ for a $65\,M_\odot$ black hole and $24\,\mathrm{min}$ for a $10^6\,M_\odot$ supermassive black hole. Further inside the unstable region, at $(a/M,\eta/M^3)=(1.90,2.62)$, the growth rate reaches $\mathrm{Im}(M\omega)\simeq1.05\times10^{-1}$, giving $\tau\simeq9.5M$, or about $3\,\mathrm{ms}$ for a $65\,M_\odot$ black hole and $47\,\mathrm{s}$ for a $10^6\,M_\odot$ supermassive black hole.
The fact that the growth rate increases as we move deeper into the unstable region suggests that the instability becomes even stronger beyond the parameter range explored here.

We also consider other multipoles to determine whether the instability is restricted to the $(\ell,m)=(2,2)$ mode. Figure~\ref{fig:S21} shows $\mathrm{Im}(M\omega)$ for the corotating $(\ell,m)=(2,1)$ gravitational mode. This mode is also unstable in part of the superspinning region, although its growth rates are smaller than those of the $(\ell,m)=(2,2)$ mode for the same black hole parameters. The axisymmetric $(\ell,m)=(2,0)$ mode, on the other hand, remains damped
throughout the explored parameter space, as shown in Fig.~\ref{fig:S20}.

\begin{figure}[h!]
\centering
\includegraphics[width=0.99 \linewidth]{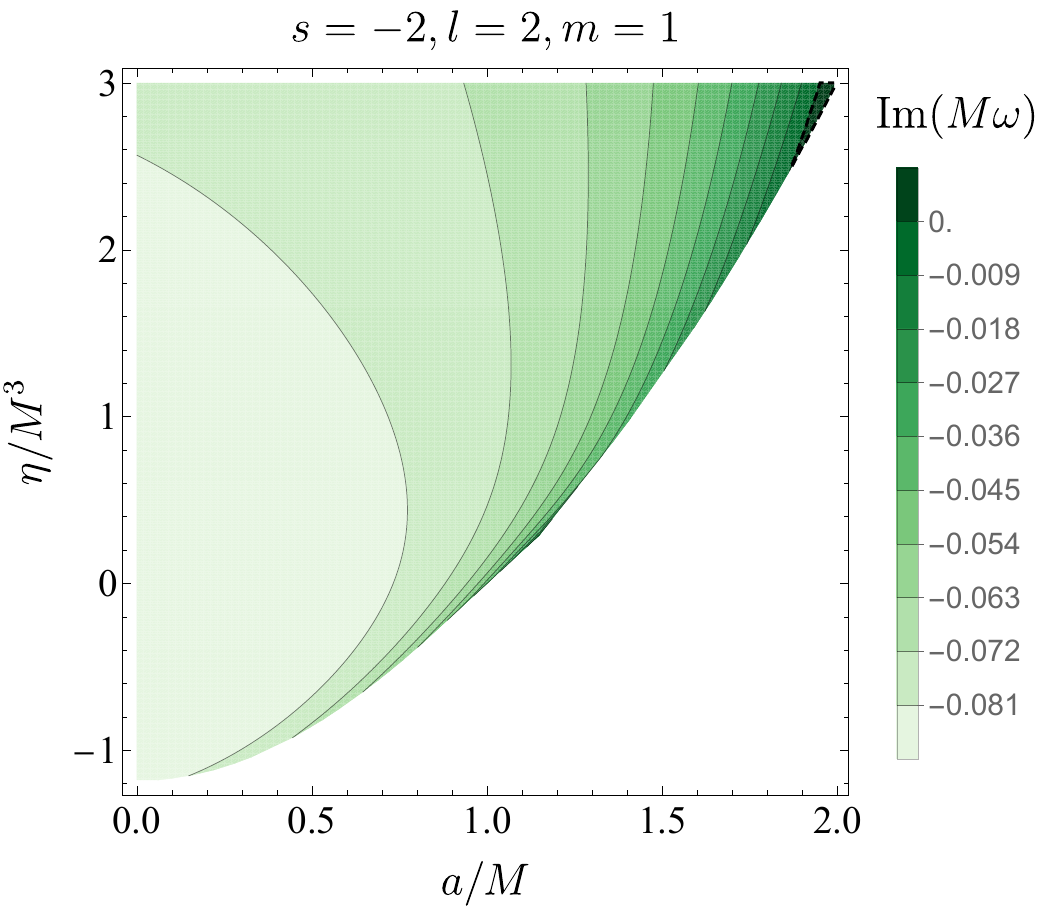}
\caption{Contour plot of $\text{Im}(M\omega)$ for the fundamental $(\ell,m)=(2,1)$ gravitational modes. The black dashed line (top right) encloses the region with $\operatorname{Im}(M\omega)>0$, where the homogeneous spin-$2$ operator admits exponentially growing modes.}

\label{fig:S21}
\end{figure}

\begin{figure}[h!]
\centering
\includegraphics[width=0.99 \linewidth]{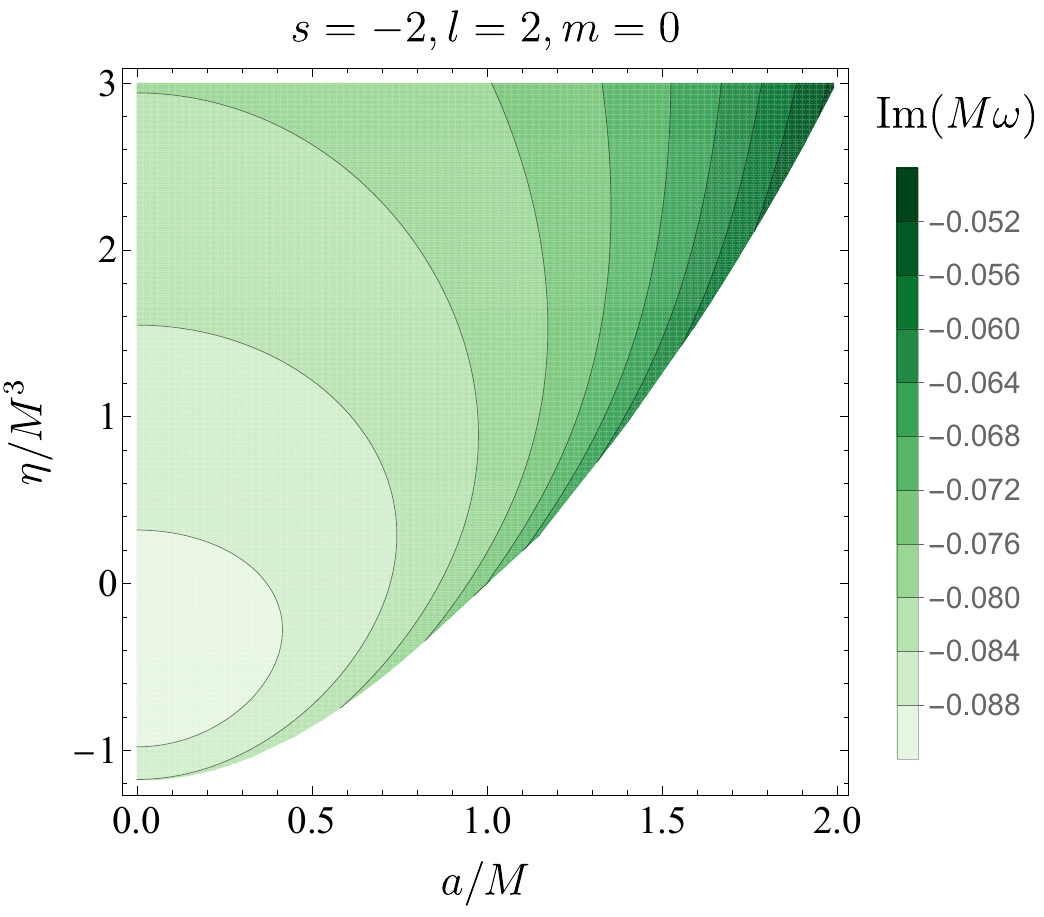}
\caption{Contour plot of $\text{Im}(M\omega)$ for  
$(\ell,m)=(2,0)$ fundamental gravitational modes.
Note that $\mathrm{Im}(M\omega)$ remains negative throughout the explored parameter space.}
\label{fig:S20}
\end{figure}

\noindent \textbf{\textit{GW250114 ringdown compatibility --}} 
We investigate the compatibility of the QNMs of the KRZ geometry with the GW250114 ringdown. 
Specifically, we identify KRZ configurations whose damped modes are consistent with the posterior obtained from a model-agnostic analysis of the GW250114 ringdown, without constructing a posterior for the KRZ parameters. 
%Specifically, we test which KRZ configurations yield damped modes compatible with the agnostic ringdown posterior, without constructing a posterior distribution for the KRZ parameters. 
To this end, we use the observational dataset~\cite{LIGOScientific:2025wao, zenodo250114}, which provides time-domain extractions of the ringdown frequencies $f_{\text{obs}}$ and damping times $\tau_{\text{obs}}$ without assuming a specific GR ringdown template.

To map the observables $(f_{\text{obs}}, \tau_{\text{obs}})$ to the dimensionless parameter space $(a/M, \eta/M^3)$, we use the remnant mass $M$, inferred using the NRSur7dq4 waveform model~\cite{LIGOScientific:2025rid}
%, for the event~\cite{LIGOScientific:2025wao} 
to set the mass scale. This choice provides a reference mass in the absence of complete numerical simulations of KRZ mergers. For decaying modes, the physical frequency $f_{\text{th}}$ and damping time $\tau_{\text{th}}$ are given by
\begin{equation}
	f_{\text{th}} = \frac{\operatorname{Re}(M\omega) c^3}{2\pi G M} \,, \quad \quad  \tau_{\text{th}} = \frac{G M}{-\operatorname{Im}(M\omega) c^3} \,. \label{eq:dimensional_restore}
\end{equation}

We account for the remnant mass uncertainty by evaluating a discrete grid of $N_M = 200$ mass values uniformly sampled over the GR-inferred $90\%$ credible interval $ [67.2 \,M_\odot, 68.9\,M_\odot]$. To determine whether the theoretical QNM predictions are consistent with the data, we fit a 2D Gaussian Kernel Density Estimate (KDE) to the posterior samples in the $(f_{\text{obs}}, \tau_{\text{obs}})$ plane, as detailed in Supp. Material~\cite{supplemental}. The $90\%$ credible boundary is set by the 10th percentile of the KDE density evaluated on the observational posterior samples. We interpolate the resulting probability density over a $500 \times 500$ regular grid, which allows us to evaluate the observational distribution at arbitrary points in the $(f_{\text{obs}},\tau_{\text{obs}})$ plane.

We then compare the observational distribution with the theoretical predictions across the KRZ parameter space. For damped modes ($\operatorname{Im}(M\omega) < 0$), a point $(a/M, \eta/M^3)$ of the parameter  space is considered compatible with the data if, for at least one mass value in the tested range, the corresponding pair $(f_{\text{th}}, \tau_{\text{th}})$ lies within the $90\%$ credible region of the observational distribution. Exponentially growing modes, with $\operatorname{Im}(M\omega)\geq0$, do not have a positive damping timescale and therefore cannot be compared with the observed damping times. %Hence, they have zero density under the observational distribution and are excluded from the allowed parameter space.

To account for the time dependence of the ringdown, we apply this procedure to three extraction windows starting at $0 M$ ($t_0$), $10 M$ ($t_{10}$), and $20 M$ ($t_{20}$) after the peak strain amplitude. The early window provides higher signal-to-noise ratio (SNR) but may contain residual nonlinearities and prompt ringdown excitations from the merger~\cite{Bhagwat:2016ntk, Giesler:2019uxc}. The late window is less affected by nonlinear effects, but has lower SNR due to signal attenuation. The intermediate window provides a balance between linearity and SNR. Fig.~\ref{fig:restriction_eta} shows the resulting regions in the $(a/M, \eta/M^3)$ plane. The constraints obtained 
from the three extraction windows are consistent with one another and include superspinning configurations. Thus, combining the QNM spectrum with the agnostic GW analysis allows us to identify regions of the KRZ parameter space whose QNM predictions are consistent with the GW250114 ringdown observations.

\begin{figure}[t]
	\centering
	\includegraphics[width=\columnwidth]{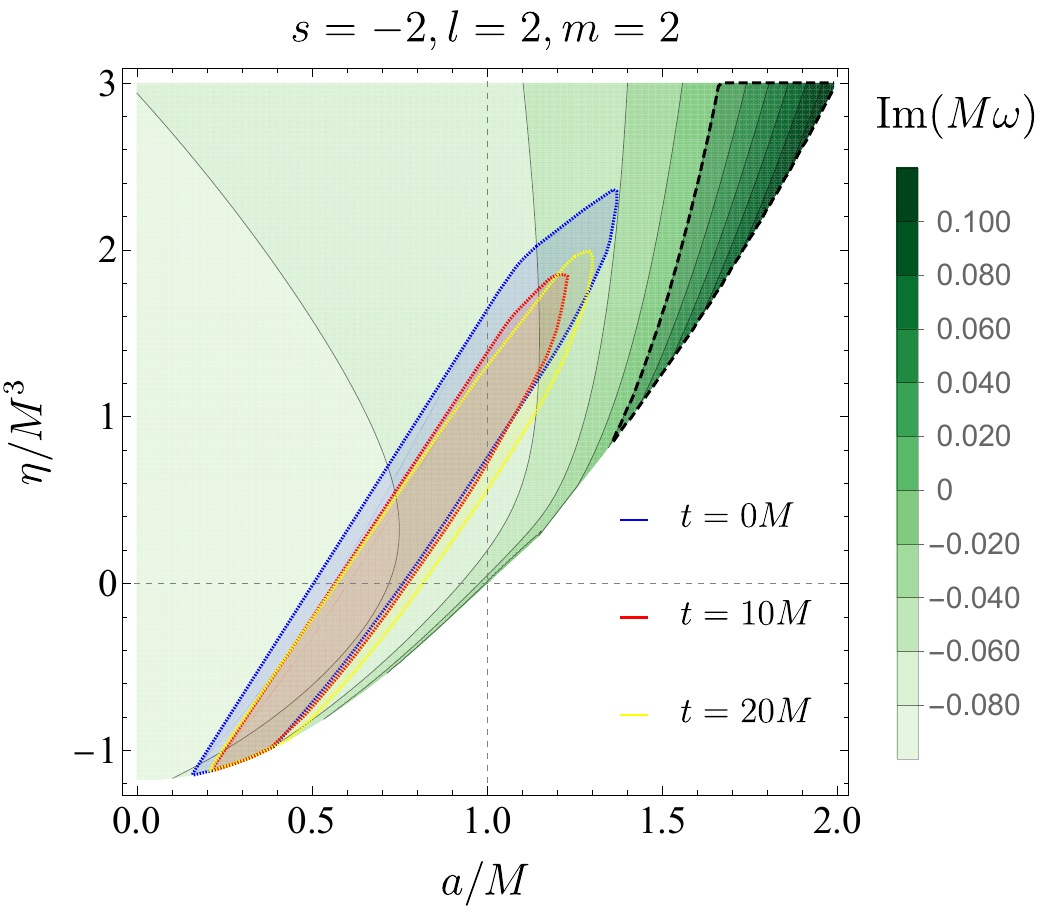}
	\caption{
    Contour plot of $\operatorname{Im}(M\omega)$ for the fundamental $(\ell,m)=(2,2)$ gravitational QNM, as in the bottom-left panel of Fig.~\ref{fig:s2es1}. The thick black dotted line encloses the region with $\operatorname{Im}(M\omega)>0$, where the modes are unstable. The blue, red, and yellow dotted contours enclose KRZ configurations whose predicted QNMs lie within the $90\%$ credible regions obtained from the agnostic time-domain analysis of the GW250114 ringdown at early ($t_0$), intermediate ($t_{10}$), and late ($t_{20}$) time bins, respectively.}
    \label{fig:restriction_eta}
\end{figure}
%%%%%%%%%%%%%%%%%%%%%%%%%%%%%%%%%%%%%%%%%%%%%%%%%%%%%%%%%%%%%%%%%%%%%%%%%%%%%%%%%%%%%%%%%%%%%%%%

\noindent \textbf{\textit{Discussion --}} In this Letter, we have identified exponentially growing modes of a homogeneous spin-weight $s=-2$ Teukolsky-like operator, subject to QNM boundary conditions, on a one-parameter KRZ black hole geometry. The instabilities arise in highly spinning and strongly deformed regimes, including superspinning ($a>M$) black hole configurations. The $(\ell,m)=(2,2)$ and $(2,1)$ branches develop positive imaginary frequencies, with growth rates reaching $\operatorname{Im}(M\omega)\simeq 0.105$, corresponding to e-folding times as short as $\tau\simeq 9.5\,GM/c^{3}$, within the explored domain. In contrast, the test electromagnetic modes examined remain damped. 
%Electromagnetic probes, such as X-ray reflection spectroscopy and shadow imaging, are therefore insensitive to the instabilities identified here.

Unlike horizonless Kerr superspinars, which lack an event horizon and whose perturbation spectra consequently depend on boundary conditions imposed at an inner surface~\cite{Pani:2010jz}, the configurations considered here possess an event horizon and are subject to standard black hole QNM boundary conditions. Furthermore, the instabilities identified here are distinct from superradiant instabilities, which arise when quasibound states become unstable and involve different boundary conditions in the asymptotic region~\cite{brito2020}. Our results thus point to a potentially distinct instability mechanism, although their physical interpretation remains uncertain. In particular, the KRZ background is not Ricci-flat, and its supporting dynamical theory is unspecified. Hence, the growing modes establish an instability of the homogeneous Teukolsky-like operator, but do not necessarily imply an instability of a self-consistent gravitational perturbation system. A key challenge for future work is to determine whether these instabilities persist in a self-consistent framework in which gravitational perturbations are coupled to the dynamical degrees of freedom supporting the non-Kerr background.

Applying our agnostic time-domain analysis to GW250114, we identify regions of the KRZ parameter space whose QNM predictions are consistent with the observed ringdown data. These results constrain the spin and deformation parameters but do not exclude significant deviations from Kerr. Given that X-ray measurements and analyses of the inspiral phase currently provide stronger constraints on deviations from Kerr~\cite{Shashank:2021giy,Abdikamalov:2021zwv,Yu:2021xen,Cardenas-Avendano:2016zml,Konoplya:2016pmh}, our aim is not to obtain the tightest possible bounds, but rather to demonstrate the potential of ringdown observations to constrain Kerr-like geometries. Future observations with third-generation detectors, such as the Einstein Telescope, and space-based detectors, such as LISA~\cite{Punturo:2010zz,LISA:2017pwj,Berti:2005ys,Baibhav:2019rsa}, are expected to yield higher SNRs, enabling more precise black hole spectroscopy~\cite{Berti:2025hly} and providing further opportunities to test the Kerr hypothesis and investigate the QNM spectra of Kerr-like geometries.

\noindent \textbf{\textit{Data Availability --}} Data supporting the findings of this work are openly available~\cite{dataKRZ}.

%%%%%%%%%%%%%%%%%%%%%%%%%%%%%%%%%%%%%%%%%%%%%%%%%%%%%%%%%%%%%%%%%%%%%%%%%%%%%%%%%%%%%%%%%%%%%%%%

\noindent \textbf{\textit{Acknowledgements --}} This research was partially financed by the Coordena\c{c}\~ao de Aperfei\c{c}oamento de Pessoal de N\'{i}vel Superior (CAPES, Brazil) - Finance Code 001. P. H. C. S. acknowledges support from the São Paulo Research Foundation (FAPESP, Brazil, grants 2022/07298-4 and 2026/14809-6). R. D. B. F. acknowledges {\it Conselho Nacional de Desenvolvimento Científico e Tecnológico} (CNPq) under Grants No. $309304/2026-1$, $407868/2025-9$ and $405749/2023-6$. M.R. acknowledges partial support from the Conselho Nacional de Desenvolvimento Científico e Tecnológico (CNPq, Brazil, grant 315991/2023-2), and from the São Paulo Research Foundation (FAPESP, Brazil, grant 2024/00923-6). The Tycho supercomputer hosted at the SCIENCE HPC center at the University of Copenhagen was used for supporting this work.
%%%%%%%%%%%%%%%%%%%%%%%%%%%%%%%%%%%%%%%%%%%%%%%%%%%%%%%%%%%%%%%%%%%%%%%%%%%%%%%%%%%%%%%%%%%%%%%%

\noindent \textbf{\textit{AI use --}} The authors used OpenAI's ChatGPT (GPT-6) to assist with checking Newman–Penrose calculations, writing and debugging Python code for the analysis leading to Fig.~5, and improving the clarity of the manuscript. All AI-assisted calculations and code were independently verified, and all textual suggestions were reviewed by the authors, who take full responsibility for the content of this work.
%The authors used GPT-6 to help check Newman–Penrose calculations, write and debug Python code, and improve the manuscript's text. All results were checked by the authors, who take full responsibility for the content of this work.

%\bibliographystyle{apsrev4-2}
\bibliography{refs}

% Supplemental Material
\clearpage
\onecolumngrid
 %\documentclass[aps,prl,onecolumn,amssymb,amsmath,a4paper,superscriptaddress,longbibliography]{revtex4-2}
% \usepackage{graphicx}
% \usepackage{amsfonts}
% \usepackage{amsmath}
% \usepackage{hyperref}
% \usepackage{xr-hyper}
% \externaldocument[main-]{main}
% \usepackage{mathrsfs}
% \usepackage{adjustbox}
% \usepackage{float} 
% \usepackage[normalem]{ulem}
% \usepackage[nameinlink]{cleveref}
% \usepackage[table,xcdraw]{xcolor}
% \usepackage[mathcal]{euscript}
\Crefname{figure}{Fig.}{Figs.}
\def\be{\begin{eqnarray}}
\def\ee{\end{eqnarray}}
\def\nn{\nonumber}

%\begin{document}

\title{Supplemental Material for: Gravitational Mode Instabilities in Kerr-like Black Holes}

\author{Pedro Henrique Croti Siqueira}
\email{pedro.croti@ufabc.edu.br}
\affiliation{Centro de Matem\'atica, Computa\c c\~ao e Cogni\c c\~ao, Universidade Federal do ABC (UFABC), 09210-170 Santo Andr\'e, S\~ao Paulo, Brazil}
\affiliation{Centro de Ci\^encias Naturais e Humanas, Universidade Federal do ABC (UFABC), 09210-170 Santo Andr\'e, S\~ao Paulo, Brazil}
\affiliation{Instituto de Física Gleb Wataghin, Universidade Estadual de Campinas, 13083-059 Campinas, São Paulo, Brazil}
\author{Rodrigo Dal Bosco Fontana}
\email{rodrigo.dalbosco@ufrgs.br}
\affiliation{Universidade Federal do Rio Grande do Sul (UFRGS), 95590-000, Tramandaí, Rio Grande do Sul, Brazil}

\author{Maur\'icio Richartz}
\email{mauricio.richartz@ufabc.edu.br}
\affiliation{Centro de Matem\'atica, Computa\c c\~ao e Cogni\c c\~ao, Universidade Federal do ABC (UFABC), 09210-170 Santo Andr\'e, S\~ao Paulo, Brazil}

\maketitle

\begin{center}
{\large\bfseries Supplemental Material}\\[1ex]
\end{center}

\setcounter{equation}{0}
\setcounter{figure}{0}
\setcounter{table}{0}
\setcounter{page}{1}

\makeatletter
\renewcommand{\theequation}{S\arabic{equation}}
\renewcommand{\thefigure}{S\arabic{figure}}
\renewcommand{\bibnumfmt}[1]{[#1]}
\renewcommand{\citenumfont}[1]{#1}

\section{I. Derivation of the Master Equation}

We consider linear gravitational perturbations of the KRZ black hole spacetime (1). Here we outline the key elements of the Newman--Penrose (NP) formalism required for deriving the perturbation equations, building on  previous works~\cite{tkk1, tkk2, tkk3, chandra1, frolov, brandeis64, newmanpenrose, yunes22, hussain22}. Throughout this section, we adopt the standard NP conventions of Chandrasekhar~\cite{chandra1}.

The deformed Kerr spacetime (1) is defined by the following null tetrad vectors
\be
\label{egp0a}
l & =  {l^\mu \partial_\mu} = & \frac{r^2+a^2}{f} {\partial_t} + {\partial_r} + \frac{a}{f} {\partial_\varphi}, \\
\label{egp0b}
n & =  {n^\mu \partial_\mu} = & \frac{r^2+a^2}{2\Sigma} {\partial_t}  - \frac{f}{2\Sigma} {\partial_r} + \frac{a}{2\Sigma} {\partial_\varphi}, \\
\label{egp0c}
m & =  {m^\mu \partial_\mu} = & \frac{ia\sin \theta}{\sqrt{2}(r+ia\cos \theta)} {\partial_t} + \frac{1}{\sqrt{2}(r+ia\cos \theta)} {\partial_\theta}  + \frac{i\csc \theta }{\sqrt{2}(r+ia\cos \theta)} {\partial_\varphi},  
\ee
with $f=r^2-2Mr + a^2 - \eta/r$.  The background spin coefficients are
\be
\nn
\kappa = \sigma = \lambda = \nu = \epsilon = 0;  \hspace{0.7cm} \rho = \frac{1}{-r+ia\cos \theta}; \hspace{0.4cm}
\nn
\beta = -\frac{\rho^* \cot \theta }{2\sqrt{2}}; \hspace{0.4cm} \mu = \frac{f \rho}{2\Sigma}; \hspace{0.4cm} \tau = \frac{-ia\sin \theta}{\sqrt{2}\Sigma};
\nn \\
\pi = \frac{\rho^2ia\sin \theta}{\sqrt{2}}; \hspace{0.4cm} \gamma = \mu +\frac{\partial_r f}{4\Sigma};   \hspace{0.4cm}  \alpha = \pi - \beta^*, 
\label{egp4}
\ee
while the nonzero projected Ricci tensor components are
\be
\nn
\Phi_{00}  & = &  \Phi_{01}=\Phi_{02}=\Phi_{10}=\Phi_{12}=\Phi_{20}=\Phi_{21}=\Phi_{22}=0;  \\
\Phi_{11}  & = &  %\frac{4r^2-4a^2 + 4f - 4r\partial_r f+\Sigma \partial_{rr}f}{8\Sigma^2}= 
- \frac{\eta(4r^2+\Sigma)}{4{r^3}\Sigma^2}; \hspace{1cm} \Lambda   %=  \frac{2-\partial_{rr}f }{12\Sigma} 
= \frac{\eta}{12r^3\Sigma}. \label{egp5}
\ee
Since the spacetime is Petrov type-D, the only nonzero background Weyl scalar is
\be
%\nn
\Psi_2 = %\frac{1}{12\Sigma}(\partial_{rr}f -2 - 12 \rho^2a^2\sin^2\theta  \nn
%+ 12ia\rho \cos \theta+12\rho^2 f +6 \rho \partial_r f ) = \\ 
%M\rho^3 + \eta \rho^3(-10r^2 + 5iar\cos \theta + a^2\cos^2 \theta)\frac{\rho^*}{6r^3}. 
 {-\frac{M}{(r - i a \cos\theta)^3} + \frac{\eta \left(-10 r^2 + 5 i a r \cos\theta + a^2 \cos^2\theta\right)}{6 r^3 (r - i a \cos\theta)^3 (r + i a \cos\theta)}}. \label{egp6}
\ee
All quantities in Eqs.~\eqref{egp4}--\eqref{egp6} are evaluated on the unperturbed background.

We next derive the equations governing linear gravitational perturbations. We begin with the relevant NP equations in their exact, nonlinear form. Two equivalent formulations are commonly used, organizing the equations into two independent systems with either $\Psi_4$ or $\Psi_0$ as the fundamental perturbation variable~\cite{tkk1, chandra1, frolov, chen24, xiong23}. The two formulations are related by the Geroch--Held--Penrose formalism, which interchanges the null tetrad directions~\cite{Geroch:1973am}.
 Here, we adopt the $\Psi_4$ formulation (the transformation from the $\Psi_0$ formulation can be found, e.g., in Ref.~\cite{chen24}). The relevant equations of motion are~\cite{guo24}
%%%
\begin{align}
\delta^* \Psi_3 - D \Psi_4 + \delta^*\Phi_{21} - \Delta \Phi_{20} = (4\varepsilon - \rho) \Psi_4 &- (4\pi + 2\alpha)\Psi_3 + 3\lambda \Psi_2 + (2\gamma - 2\gamma^* + \mu^*) \Phi_{20} \nonumber \\
&\quad + 2(\tau^* - \alpha)\Phi_{21} + 2\lambda \Phi_{11} - 2\nu \Phi_{10} - \sigma^* \Phi_{22}, \label{egp1} \\[1.5ex]
\Delta \Psi_3 - \delta \Psi_4 + \delta^* \Phi_{22} - \Delta \Phi_{21} = (4\beta - \tau) \Psi_4 &- (4\mu + 2\gamma)\Psi_3 + 3\nu \Psi_2 + (\tau^* - 2\beta^* - 2\alpha) \Phi_{22} \nonumber \\
&\quad + 2(\gamma + \mu^*)\Phi_{21} + 2\lambda \Phi_{12} - 2\nu \Phi_{11} - \nu^* \Phi_{20}, \label{egp2} \\[1.5ex]
\Delta \lambda + (\mu + \mu^* + 3\gamma - &\gamma^*)\lambda  {-} \delta^* \nu - (\pi - \tau^* + \beta^* + 3\alpha) \nu + \Psi_4 = 0, \label{egp3}
\end{align}
where $D = l^\mu \nabla_\mu$, $\Delta = n^\mu \nabla_\mu$ and $\delta = m^\mu \nabla_\mu$. Being exact, these equations contain both background and perturbative contributions. We now linearize them about the KRZ background specified above, retaining only terms of first order in the perturbations. We denote perturbative quantities by a superscript $(P)$, while all quantities without this superscript refer to the background. The resulting linearized equations are
{\small
\begin{align}
(\delta^* + 4\pi +2\alpha )\Psi_3^{(P)} -(D -\rho )\Psi_4^{(P)}  {-}     (3\Psi_2+2\Phi_{11})\lambda^{(P)} =
(- \delta^* + 2\tau^* -2\alpha )\Phi_{21}^{(P)}  {+} (\Delta  {+} 2\gamma  {-} 2\gamma^*  {+} \mu^* )\Phi_{20}^{(P)} 
\label{egp7}
\\
(\Delta+4\mu + 2\gamma ) \Psi_3^{(P)} - ( \delta -\tau +4\beta ) \Psi_4^{(P)}  {-} (3\Psi_2  {-} 2\Phi_{11})\nu^{(P)}  = 
(- \delta^* +\tau^* -2\beta^* - 2\alpha)\Phi_{22}^{(P)}   {+} ( \Delta  {+} 2\gamma  {+} 2\mu^* )\Phi_{21}^{(P)} \label{egp8}\\
\Delta \lambda^{(P)}  + (\mu + \mu^* + 3\gamma -\gamma^* )\lambda^{(P)}  {-} \delta^* \nu^{(P)} 
  {-} (\pi -\tau^* + \beta^* + 3\alpha ) \nu^{(P)} + \Psi_4^{(P)} = 0,
\label{egp9}
\end{align}}
where terms of quadratic and higher order in the perturbations have been neglected. From now on, we suppress the superscript $(P)$ on first-order perturbative quantities to simplify the notation.

To decouple the system and isolate $\Psi_4$, we apply appropriate differential operators to Eqs.~\eqref{egp7} and \eqref{egp8} and subtract the resulting equations. Because the NP directional derivative operators do not commute, the elimination of $\Psi_3$ requires the use of the following NP relation,
\be
(\Delta + (p+1)\gamma - \gamma^* - q\mu + \mu^*)(\delta^* + p\alpha -q\pi ) = 
(\delta^* + (p+1)\alpha + \beta^* - \tau^* - q\pi ) (\Delta +p\gamma -q\mu). 
\label{egp10}
\ee
Specifically, we fix $q=-2p=-4$ and define the operators $\widehat{\mathcal{O}}_\Delta$, $\widehat{\mathcal{O}}_{\delta^*} $, $\widehat{\mathcal{O}}_{\delta}$, $\widehat{\mathcal{O}}_{D}$ through
\be
\label{egp11}
\widehat{\mathcal{O}}_\Delta & = & \Delta + 3\gamma - \gamma^* +4\mu + \mu^*, \\
\label{egp12}
\widehat{\mathcal{O}}_{\delta^*} & = & \delta^* + 3\alpha + \beta^* - \tau^* +4\pi , \\
\label{egp13}
\widehat{\mathcal{O}}_{\delta} & = & \delta + 4\beta - \tau  , \\
\label{egp14}
\widehat{\mathcal{O}}_{D} & = & D + 4\epsilon - \rho. 
\ee
Next, we apply $\widehat{\mathcal{O}}_\Delta$ to Eq.~\eqref{egp7} and $\widehat{\mathcal{O}}_{\delta^*}$ to Eq.~\eqref{egp8}, and subtract the resulting expressions to obtain
\be
\mathcal{T} +\widehat{\mathcal{O}}\Psi_4 + (3\Psi_2+2\Phi_{11})\widehat{\mathcal{O}}_\Delta \lambda + \lambda \Delta (3\Psi_2+2\Phi_{11}) %\\
\label{egp15}
-(3\Psi_2-2\Phi_{11})\widehat{\mathcal{O}}_{\delta^*} \nu - \nu \delta^* (3\Psi_2-2\Phi_{11}) =0, 
\ee
where $\widehat{\mathcal{O}}  \equiv  \widehat{\mathcal{O}}_{\Delta}\widehat{\mathcal{O}}_{D} - \widehat{\mathcal{O}}_{\delta^*}\widehat{\mathcal{O}}_{\delta}$, and $\mathcal{T}$ is given by
\be
\! \! \! \!  \! \!  \mathcal{T} \! = \!  {-}\widehat{\mathcal{O}}_{\delta^*} \Big[(\Delta + 2\mu^* +2\gamma )\Phi_{21} - (\delta^* - \tau^*+2\alpha +2\beta^* )\Phi_{22} \Big] \!  %\\
 - \widehat{\mathcal{O}}_\Delta  \Big[  (\delta^* - 2\tau^* +2\alpha )\Phi_{21} -  (\Delta + \mu^* +2\gamma -2\gamma^* )\Phi_{20} \Big]\!.
\label{egp16}
\ee

In the pure Kerr geometry, the background identities $(\Delta + 3\mu)\Psi_2 = 0$ and $(\delta^* + 3\pi)\Psi_2 = 0$ allow Eqs.~\eqref{egp9} and \eqref{egp15} to be directly combined into $\mathcal{T} + (\widehat{\mathcal{O}} - 3\Psi_2)\Psi_4 = 0$, as shown in Ref.~\cite{tkk1}. For the deformed rotating black hole considered here, we adopt the approach of Refs.~\cite{nakajima25, guo24, jing24}. Specifically, using the identities
\be
\Delta(3\Psi_2-2\Phi_{11}) =&-3\mu(3\Psi_2+2\Phi_{11}) +4(\mu+\mu^*)\Phi_{11},
\\
\delta^*(3\Psi_2+2\Phi_{11}) =&-3\pi(3\Psi_2-2\Phi_{11}) -4(\pi-\tau^*)\Phi_{11},
\ee
we add $(-3\Psi_2-6\Lambda )$ times Eq.~\eqref{egp9} to \eqref{egp15} thereby obtaining
\be
 ( \widehat{\mathcal{O}} - 3\Psi_2 - 6\Lambda ) \Psi_4  + \Xi + \mathcal{T}  =0,
\label{egp20}
\ee
with
\be
\Xi={}& 2(\Phi_{11}-3\Lambda)(\widehat{\mathcal{O}}_\Delta-3\mu)\lambda + 4 \lambda(\Delta+\mu+\mu^*)\Phi_{11} + 2(\Phi_{11}+3\Lambda)(\widehat{\mathcal{O}}_{\delta^*}-3\pi)\nu + 4\nu(\delta^*+\pi-\tau^*)\Phi_{11}.
\ee
A separated gravitational perturbation equation can be derived through a type-I tetrad transformation~\cite{chandra1, frolov, yunes22, hussain22, guo24, jing24} of order $(P)$, namely $m \rightarrow m+b^{(P)}l \equiv  m+bl$. Taking into account the induced changes in every spin coefficient, Ricci term, directional derivative and Weyl scalar, the only modifications that contribute to (\ref{egp20}) are 
\be
\nn
\lambda & \rightarrow & \lambda + (2\alpha + \pi + \delta^*) b^*, \\
\nn
\nu & \rightarrow & \nu  + (2\gamma + \mu + \Delta ) b^*, \\
\label{egp24}
\Phi_{21} & \rightarrow & \Phi_{21} + 2\Phi_{11} b^*.
\ee
The changes induced by the tetrad rotation modify both $\Xi$ and $\mathcal{T}$ of Eq.~\eqref{egp20}, which can be 
rewritten as
\be
\Xi_g&=&\Xi +4\Big\{(\delta^*+3\alpha+\beta^*-2\tau^*+2\pi)\left[\Phi_{11}(\Delta+2\gamma+\mu)b^*\right]+\big[(\Delta+\mu+\mu^*)\Phi_{11}\big](\delta^*+2\alpha+\pi)b^*\Big\}, \\
\mathcal{T}_g&=&\mathcal{T}-4\Big\{(\delta^*+3\alpha+\beta^*-2\tau^*+2\pi)\left[\Phi_{11}(\Delta+2\gamma+\mu)b^*\right]+\big[(\Delta+\mu+\mu^*)\Phi_{11}\big](\delta^*+2\alpha+\pi)b^* + \mathcal{R}b^*\Big\}, \ \ \ \ \  
\ee
where 
\be
\mathcal{R} = (\delta^*+\alpha+\beta^*-\tau^*+3\pi)\left[(\Delta+\mu+\mu^*)\Phi_{11}\right]-(\Delta+\gamma-\gamma^*+3\mu+\mu^*)\left[(\tau^*+2\pi)\Phi_{11}\right].
\ee
%with $\mathcal{T}=\mathcal{T}_g$. 
Now, by choosing $b$ appropriately, we can impose the separable gauge $\Xi_g=0$, thus simplifying the perturbation equation. This gauge condition is associated with the freedom to perform infinitesimal type-I null tetrad rotations, which leaves $\Psi_4$ invariant at linear order~\cite{chandra1,guo24}.

Since the KRZ geometry is not Ricci-flat, obtaining it as a solution of the gravitational field equations requires either additional matter sources within GR or modifications to the gravitational theory itself. In either case, the underlying theory may introduce additional degrees of freedom whose perturbations can couple to the gravitational sector. Such couplings, however, cannot be inferred from the background geometry alone. In the gauge discussed above, these contributions enter through the effective source $\mathcal{T}_g$, which involves perturbations of the Ricci tensor, see Eq.~\eqref{egp16}. In this work, we neglect possible couplings to additional degrees of freedom by setting $\mathcal{T}_g=0$, thus restricting our analysis to the homogeneous equation
\begin{equation}
(\widehat{\mathcal{O}}-3\Psi_2-6\Lambda)\Psi_4=0.
\end{equation}
The left hand side of the equation above defines the gravitational operator studied here.

Finally, by adopting the transformation $\Psi_4 = \rho^4 \psi_{-2}$,
the perturbation equation takes the explicit form
\be
\nn
\left( \frac{(r^2+a^2)^2}{f}-a^2\sin^2\theta \right) \frac{\partial^2 \psi_{-2}}{\partial t^2}   - f \frac{\partial^2 \psi_{-2}}{\partial r^2}    %\\ 
\nn
+\left( \frac{2(r^2+a^2)\partial_r f}{f} -8r-4ia\cos \theta \right) \frac{\partial \psi_{-2}}{\partial t} 
+\partial_r f \frac{\partial \psi_{-2}}{\partial r}  \\
\nn
- \frac{\partial^2 \psi_{-2}}{\partial \theta^2}   + \left(\frac{2a(r^2+a^2)}{f}-2a\right) \frac{\partial^2\psi_{-2}}{\partial t\,\partial\phi} + \left( \frac{a^2}{f}-\frac{1}{\sin^2 \theta } \right) \frac{\partial^2 \psi_{-2}}{\partial \varphi^2}   - \cot \theta \frac{\partial \psi_{-2}}{\partial \theta}  %\\
\nn \\
+\left( 4i \cot \theta \csc \theta  + \frac{2a\partial_r f}{f} \right) \frac{\partial \psi_{-2}}{\partial \varphi}  + (4\cot^2 \theta +2) \psi_{-2} =0.
\label{egp24b}
\ee
The resulting equation is separable under the standard mode decomposition 
\be
\label{egp25}
\psi_{-2} = e^{-i\omega t}e^{i m \varphi} R (r) S (\theta ).
\ee
The separated angular equation is the spin-weighted spheroidal wave equation~\cite{Berti:2005gp}. The separated radial equation takes the standard form given in Eq.~(3). We note that the corresponding perturbation equations for electromagnetic ($s=\pm1$) and scalar ($s=0$) fields have been derived in Refs.~\cite{Liberati1, Liberati2, Siqueira:2022tbc}, and also satisfy Eq.~(3).

\section{II. Numerical Implementation of the continued fraction method} 

The QNM frequencies associated with Eq.~(3) are computed using Leaver's continued fraction method~\cite{Leaver:1985ax}. Imposing the QNM boundary conditions and factoring out the corresponding asymptotic behavior at the singular points yields the following \textit{ansatz} for the radial function:
\begin{align}
R\left( r\right)&=e^{i\omega r}\left( r-r_{0}\right) ^{-s-i \sigma_{0}} (r-r_{2})^{i \sigma_{2}} %\nonumber %\\
%\times & 
\left( r-r_{1}\right)^{i\sigma_0 -i\sigma_2 +\chi -1 -s} \sum ^{\infty }_{n=0}a_{n}\left( \frac{r-r_{0}}{r-r_{1}}\right) ^{n}, \label{eq:radial_solution_KZ}
\end{align}
where
\begin{equation}
\sigma_i = \frac{r_i[(r_i^2+a^2)\omega-am]}{\prod_{j\neq i}(r_i-r_j)} \qquad \text{and} \qquad \chi=2iM\omega.
\end{equation}
Recall that the event horizon $r_0$ is the largest positive real root of $\Delta$, see Eq.~(2b). For the remaining roots, we adopt the convention $r_1 \geq r_2$ if both are real, and $\operatorname{Im}(r_1) > \operatorname{Im}(r_2)$ if they are complex conjugates. Substituting the expansion above into the master equation (3) yields a five-term recurrence relation for the series coefficients $a_n$. The explicit expressions for the recurrence coefficients are provided in Ref.~\cite{dataKRZ}. Following standard numerical techniques~\cite{Siqueira:2022tbc}, this system is algebraically reduced via Gaussian elimination~\cite{Leaver:1990zz} to a three-term recurrence relation, 
\begin{equation}\label{eq:3-terms}
    \alpha_{n}a_{n+1} + \beta_{n}a_{n} + \gamma_{n}a_{n-1} = 0, \qquad n \geq 1,
\end{equation}
where the coefficients $\alpha_n, \beta_n$, and $\gamma_n$ are functions of the spacetime parameters $(M, a, \eta)$, the perturbation parameters $(s, \ell, m)$, and the frequency $\omega$.

Convergence of the series at spatial infinity is guaranteed if and only if the eigenfrequency $\omega$ satisfies the associated continued fraction equation,
\begin{equation} \label{cfequ}
    0 = \beta_{0} - \frac{\alpha_{0} \gamma_{1}}{\beta_{1} -}\frac{\alpha_{1} \gamma_{2}}{\beta_{2} -}\frac{\alpha_{2} \gamma_{3}}{\beta_{3} -} \dots \, .
\end{equation}
Efficient convergence is achieved through Nollert's improvement scheme~\cite{Nollert:1993zz}. By analyzing the recurrence relation generated by Eq.~\eqref{eq:radial_solution_KZ}, we derive the large-$n$ asymptotic behavior required for Nollert's improvement:
\begin{align}
\frac{a_{n+1}}{a_n} =1-\frac{(1-i) \left(r_0-r_1\right){}^2 \sqrt{\omega }}{\sqrt{n} \sqrt{\left(r_0-r_1\right){}^3}} +  \frac{ 6 r_0 r_1 -3 r_0^2 + r_1 (-5 r_1 + 2 r_2)+4 (r_0 - r_1)^2 s -4 i \left(r_0-r_1\right){}^2 \left(2 r_0+r_2\right) \omega}{4 n \left(r_0-r_1\right){}^2} , \label{nolimp1}
\end{align}
plus corrections of order $\mathcal{O}(n^{-3/2})$. 
As discussed in Ref.~\cite{Siqueira:2022tbc}, the method converges in the region delimited by the yellow lines in Fig.~1, where the inequalities
\begin{equation}
\left|\frac{r_0}{r_1}\right| > 1, 
\qquad 
\left|\frac{r_2-r_0}{r_2-r_1}\right| > 1,
\end{equation}
are both satisfied.

Larger coverage of the parameter space requires the alternative \textit{ansatz}
\begin{align}
R\left( r\right)&=e^{i \omega r}\left( r-r_{0}\right) ^{-s -i \sigma_{0}}(r-r_{1})^{-s-i \sigma_{1}}(r-r_{2})^{-s-i \sigma_{2}}
r^{-1+s+2\chi} \sum ^{\infty }_{n=0}b_{n}\left( \frac{r-r_{0}}{r}\right) ^{n}, \label{eq:radial_solution_KZ2}
\end{align}
with the corresponding asymptotic expansion for the coefficients $b_n$ given by:
\begin{align}
\frac{b_{n+1}}{b_n} & = 1-\frac{(1-i) \sqrt{r_0} \sqrt{\omega }}{\sqrt{n}} + 
 \frac{-2 i r_0 \omega -i r_1 \omega -i r_2 \omega -\frac{r_1 r_2}{2 r_0^2}+s-\frac{3}{4}}{n}, \label{nolimp2}
\end{align}
plus corrections of order $\mathcal{O}(n^{-3/2})$. This alternative \textit{ansatz} converges in the region bounded by the white lines in Fig.~1, where the inequalities
\begin{equation}
\left|\frac{r_1-r_0}{r_1}\right| > 1, 
\qquad 
\left|\frac{r_2-r_0}{r_2}\right| > 1,
\end{equation}
are both satisfied, see Ref.~\cite{Siqueira:2022tbc}. The parameter space we explore in this work corresponds to pairs $(a/M,\eta/M^3)$ for which at least one of the above approaches is applicable.

To determine the eigenfrequency $\omega$ we truncate the continued fraction at $N$ terms and solve the equation with Nollert's improvement using a root finding algorithm (we refer to the corresponding solution as $\omega_{(N)}$).
 We seek the roots of the truncated continued fraction equation, starting with a truncation order of $N=100$ and increasing it in steps of $100$. 
The iteration stops when the relative errors for both the real and imaginary parts of the frequency fall below a threshold of $\epsilon = 10^{-10}$. Specifically, writing the complex frequency as $\omega = \omega_R + i \omega_I$, we define the relative convergence for its real and imaginary parts at the $N$-th iteration as
\begin{equation}
\delta(\omega_R) = \left| \frac{\omega_{R}^{(N)} - \omega_{R}^{(N-100)}}{\omega_{R}^{(N)}} \right|, \quad \delta(\omega_I) = \left| \frac{\omega_{I}^{(N)} - \omega_{I}^{(N-100)}}{\omega_{I}^{(N)}} \right| .
\end{equation}
The iterative process is terminated when both components satisfy the precision threshold, i.e.,
\begin{equation}
\max \left[ \delta(\omega_R), \delta(\omega_I) \right] < \epsilon.
\end{equation}

To explore the parameter space $(a/M, \eta/M^3)$, we employ the following strategy: for a given mode $(s, \ell, m)$, we sweep the $(a/M, \eta/M^3)$ plane in increments of $0.01$. The calculation begins at the Schwarzschild configuration $a=\eta=0$, using its fundamental frequency as the initial seed for the root-finding algorithm. At every subsequent step, the previous solution serves as the starting guess for the next point. We first step upward along the vertical axis from $(a/M, \eta/M^3)=(0,0)$ to $(0,3)$, and then rightward along the $\eta/M^3 = 3$ line up to $(2,3)$. From each point on this top line, we step downward at fixed $a/M$ until covering the full convergence domain shown in Fig.~1.

\section{III. Numerical validation and the Dudley--Finley approximation}
\label{sec:benchmarks}

The Kerr limit provides a direct validation of our numerical implementation of the continued fraction method. Setting $\eta=0$, Eq.~(3) reduces to the standard Teukolsky radial equation. In Table~\ref{tab:validation}, we compare the fundamental QNM frequencies obtained with our continued fraction method code (denoted $\omega_{\rm num}$) against reference Kerr spectra for electromagnetic ($s=-1$) and gravitational ($s=-2$) perturbations (denoted $\omega_{\rm ref}$) obtained in Refs.~\cite{Leaver:1985ax,Berti:2009kk}.

\begin{table}[h!]
\caption{Validation of our continued fraction method implementation in the Kerr limit ($\eta=0$). The fundamental QNM frequencies are compared with reference Kerr spectra~\cite{Leaver:1985ax,Berti:2009kk}.
}
\label{tab:validation}
\begin{ruledtabular}
\begin{tabular}{ccccc}
$s$ & $(\ell,m)$ & $a/M$ & $M\omega_{\mathrm{num}}$ & $M\omega_{\mathrm{ref}}$ \\
\colrule
$-1$ & $(1,1)$ & $0.00$ & $0.2482632642 - 0.0924877180i$ & $0.2482632642 - 0.09248771795i$ \\
$-1$ & $(1,1)$ & $0.50$ & $0.2940909806 - 0.0876765068i$ & $0.2940909805 - 0.08767650681i$ \\
$-1$ & $(1,1)$ & $0.99$ & $0.4633987711 - 0.0312924130i$ & $0.4633987705 - 0.03129241302i$ \\
\colrule
$-2$ & $(2,2)$ & $0.00$ & $0.3736716844 - 0.0889623157i$ & $0.3736716844 - 0.08896231569i $ \\
$-2$ & $(2,2)$ & $0.50$ & $0.4641230260 - 0.0856388350i$ & $0.4641230260 - 0.08563883498i$ \\
$-2$ & $(2,2)$ & $0.99$ & $0.8708926587 - 0.0293904242i$ & $0.8708926587 - 0.02939042421i$ \\
\end{tabular}
\end{ruledtabular}
\end{table}

Beyond providing an exact benchmark in the Kerr limit, the radial operator considered here belongs to a broader class of separable Teukolsky-like equations that includes the Dudley--Finley approximation for Kerr--Newman black holes~\cite{Dudley:1977zz,Dudley:1978vd}. To explore this connection, we replace the radial function $\Delta(r)$ in Eq.~(3) by its Kerr--Newman counterpart
\begin{equation}
\Delta_{\rm KN} = r^2-2Mr+a^2+Q^2.
\end{equation}
The resulting radial equation is
\begin{equation}
\Delta_{\rm KN}^{-s} \frac{d}{dr} \left( \Delta_{\rm KN}^{s+1} \frac{dR_s}{dr} \right) + \left[ \frac{K^2-isK\Delta_{\rm KN}'}{\Delta_{\rm KN}} + 4is\omega r + 2am\omega - a^2\omega^2 - \Lambda_{\ell m} \right] R_s = 0,
\end{equation}
which coincides with the Dudley--Finley radial equation. This correspondence extends to the unseparated homogeneous gravitational equation \eqref{egp24b}, with $f(r)$ replaced by $\Delta_{\mathrm{KN}}$. In the Newman–Penrose formalism for the Kerr–Newman spacetime, imposing the separable gauge $\Xi_g=0$ reduces the effective source to $T_g=T+\Xi$. Setting $T_g=0$ then yields the Dudley–Finley equation, which neglects the coupling between gravitational and electromagnetic perturbations, allowing them to be described by decoupled Teukolsky-like equations. The full linearized Einstein--Maxwell equations, by contrast, describe coupled gravitoelectromagnetic perturbations whose QNM frequencies, in general, differ from those of the decoupled approximation~\cite{Berti:2005eb,Saha:2025nsg,Destounis:2026ljd}. 

As a numerical benchmark, we consider $M=1$, $a=0.2$, $Q=0.4$, and $(s,\ell,m)=(-2,2,2)$. Our continued fraction method  implementation yields
\begin{equation}
M\omega_{\rm DF} = 0.418807286633 - 0.089540112621i
\end{equation}
for the fundamental QNM, in agreement with the Dudley--Finley result reported in Ref.~\cite{Berti:2005eb}. For the same parameters, the corresponding gravitational-led mode of the full linearized Einstein--Maxwell system has frequency
\begin{equation}
M\omega_{\rm KN} = 0.409918268 - 0.088613263i
\end{equation}
\cite{Destounis:2026ljd}. The relative differences are approximately $2.17\%$ in the real part and $1.05\%$ in the magnitude of the imaginary part.
%This illustrates the accuracy that an approximate separable framework can achieve for a particular coupled system. 

\section{IV. Mapping Observational Data to the Parameter Space}
\label{app:hdi_validation}

\begin{figure}[htpb]
    \centering
\includegraphics[width=0.5\linewidth]{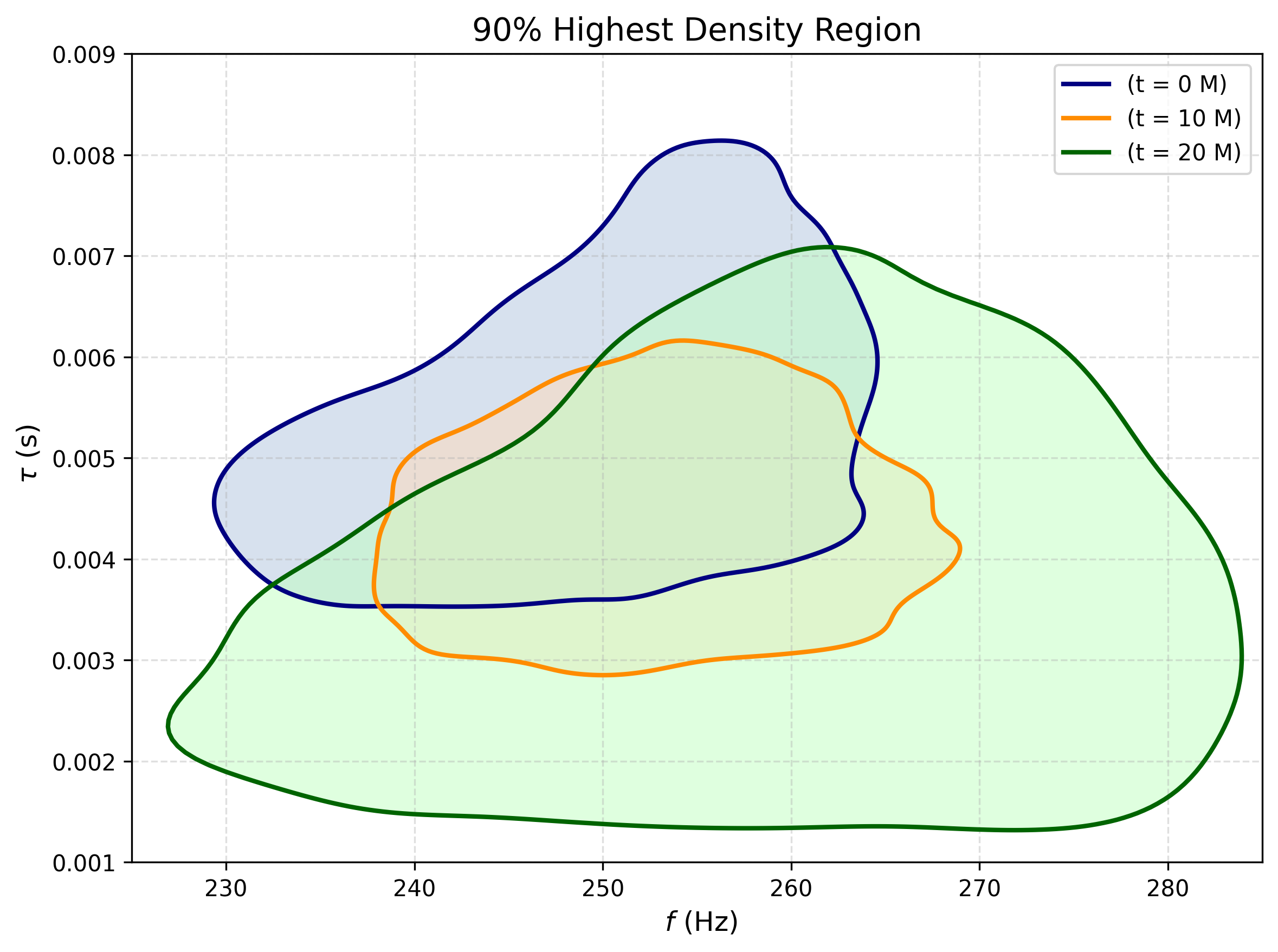}
    \caption{The $90\%$ HDR contours ($\mathcal{R}_{90}$) in the $(f,\tau)$ plane reconstructed via 2D Gaussian KDE for the three extraction windows ($t_0$, $t_{10}$, and $t_{20}$) of the GW250114 ringdown.}
    \label{fig:posterior_kde}
\end{figure}

In this section, we describe how the theoretical QNM predictions are compared with the agnostic time-domain ringdown posterior of GW250114 ~\cite{LIGOScientific:2025wao,zenodo250114}. Let $\{f_i, \tau_i\}_{i=1}^N$ denote the set of posterior samples, where $f_i$ is the ringdown frequency and $\tau_i$ is the damping time. 
We estimate the joint posterior density $\hat{p}(f,\tau\mid d)$ using a two-dimensional Gaussian kernel density estimator (KDE),
\begin{equation}
\hat{p}(f,\tau\mid d) = \frac{1}{N} \sum_{i=1}^N K_{\mathbf{H}} \left( \begin{bmatrix} f-f_i \\ \tau-\tau_i \end{bmatrix} \right),
\end{equation}
where $K_{\mathbf{H}}$ is a bivariate Gaussian kernel with bandwidth matrix $\mathbf{H}$ determined according to Scott's rule~\cite{Scott:1992}.
To construct the $90\%$ credible region, defined as the $90\%$ highest density region (HDR) of the posterior~\cite{Hyndman:1996}, we evaluate the KDE at each posterior sample,
\begin{equation}
z_i = \hat{p}(f_i,\tau_i\mid d),
\end{equation}
and set the density threshold $p_{\mathrm{th}}$ to the $10$th percentile of the set $\{z_i\}_{i=1}^N$,
\begin{equation}
p_{\mathrm{th}} = P_{10} \left( \left\{ z_i \right\}_{i=1}^N \right),
\end{equation}
where $P_{10}$ denotes the $10$th percentile. The resulting $90\%$ HDR is therefore defined as
\begin{equation}
\mathcal{R}_{90} = \left\{ (f,\tau): \hat{p}(f,\tau\mid d) \geq p_{\mathrm{th}} \right\},
\end{equation}
which retains approximately $90\%$ of the posterior samples~\cite{Hyndman:1996}. 
Fig.~\ref{fig:posterior_kde} displays the $\mathcal{R}_{90}$ contours for each extraction window after the peak strain. The KDE is evaluated on a regular Cartesian grid $(f_{\mathrm{grid}},\tau_{\mathrm{grid}})$ containing $500\times500$ points spanning the interpolation domain in both dimensions. A regular-grid interpolator $\mathcal{I}(f,\tau)$ is then constructed from this representation using the \textsc{SciPy} framework~\cite{2020SciPy-NMeth}, with points outside the domain assigned zero density.

The remnant-mass interval in the detector-frame $[M_{\text{min}}, M_{\text{max}}] = [67.2\,M_\odot, 68.9\,M_\odot]$ is discretized into $N_M = 200$ linearly spaced values,
\begin{equation}
    M_j = M_{\text{min}} + \frac{j-1}{N_M - 1}(M_{\text{max}} - M_{\text{min}}), \quad j \in \{1, \dots, N_M\}.
\end{equation}
For each theoretical configuration $(a/M,\eta/M^3)$, we use the corresponding dimensionless gravitational QNM frequency, $M\omega $, and convert its real and imaginary parts into dimensional predictions over the mass array $\{M_j\}$. Using $f = \operatorname{Re}(\omega)/(2\pi)$, the dimensional predictions for damped modes ($\operatorname{Im}(M\omega)<0$) are
\begin{equation}
f_{\mathrm{th}}(M_j) = \frac{c^3}{2\pi G} \frac{\operatorname{Re}(M\omega)}{M_j}, \qquad \tau_{\mathrm{th}}(M_j) = \frac{G}{c^3} \frac{M_j}{-\operatorname{Im}(M\omega)}.
\end{equation}
Configurations with $\operatorname{Im}(M\omega)\geq 0$ do not have a positive damping time and are therefore not included in the comparison with the observational data. A configuration $(a/M,\eta/M^3)$ is considered consistent with the $90\%$ HDR if there exists at least one sampled mass value $M_j$ for which
\begin{equation}
\mathcal{I} \left( f_{\mathrm{th}}(M_j), \tau_{\mathrm{th}}(M_j) \right) \geq p_{\mathrm{th}}.
\end{equation}

Although the 90\% HDR is obtained from the posterior in the $(f,\tau)$ plane, our procedure checks whether the theoretical QNM prediction falls within this region for at least one remnant mass in the adopted interval. This procedure yields regions in the KRZ parameter space based on the observational 90\% HDR, without assigning posterior probabilities to $(a/M,\eta/M^3)$ or marginalizing over the remnant mass. 

%\bibliography{refs}

%\end{document} 

\end{document}